\documentclass[conference]{IEEEtran}
\IEEEoverridecommandlockouts

\usepackage{listings}
\usepackage[frozencache,cachedir=.]{minted}
\usepackage{array}
\usepackage{tabularx}
\usepackage{float}

\usepackage{algorithmic}
\usepackage{graphicx}
\usepackage{textcomp}
\usepackage{xcolor}
\usepackage[hidelinks]{hyperref}

\usepackage[utf8]{inputenc}
\usepackage{url}
\usepackage{minted}
\usepackage{booktabs}
\usepackage{longtable}
\usepackage{xspace}
\usepackage[most]{tcolorbox}
\setminted{numbersep=5pt, xleftmargin=12pt}

\usepackage{multirow}

\AtBeginDocument{%
  \providecommand\BibTeX{{%
    \normalfont B\kern-0.5em{\scshape i\kern-0.25em b}\kern-0.8em\TeX}}}

\begin{document}

\title{Wasmizer: Curating WebAssembly-driven Projects on GitHub
}

\author{
\IEEEauthorblockN{Alexander Nicholson}
\IEEEauthorblockA{University of Auckland\\New Zealand}
\and
\IEEEauthorblockN{Quentin Stiévenart}
\IEEEauthorblockA{Université du Québec à Montréal\\Canada}
\and
\IEEEauthorblockN{Arash Mazidi}
\IEEEauthorblockA{TU Clausthal\\ Germany}
\and
\IEEEauthorblockN{Mohammad Ghafari}
\IEEEauthorblockA{TU Clausthal\\Germany}
}

\maketitle
\thispagestyle{plain}
\pagestyle{plain}

\begin{abstract}

WebAssembly has attracted great attention as a portable compilation target for programming languages.
To facilitate in-depth studies about this technology, 
we have deployed Wasmizer, a tool that regularly mines GitHub projects and makes an up-to-date dataset of WebAssembly sources and their binaries publicly available.
Presently, we have collected 2\,540 C and C++ projects that are highly-related to WebAssembly, and built a dataset of 8\,915 binaries that are linked to their source projects. 
To demonstrate an application of this dataset, we have investigated the presence of eight WebAssembly compilation smells in the wild.

\end{abstract} 


\pagenumbering{arabic}

\section{Introduction}

WebAssembly is a standard for portable binary code that aims to bring a safer, faster, and more portable format than JavaScript to the web. It allows for programs written in high-level languages such as C, C++, and Rust to be cross compiled to WebAssembly and run in a web environment.
WebAssembly is not limited to the web though, and it can be used in a number of host environments, e.g., it is possible to port C applications using the WebAssembly System Interface and run them as regular desktop applications.\footnote{\url{https://wasi.dev/}}

There are several tools for analyzing WebAssembly programs~\cite{Brito2022,DBLP:conf/icse/StievenartBR22,DBLP:conf/sp/RomanoLP022,DBLP:conf/icse/RomanoW20,Stievenart2020,Lehmann2019}.
However, tool builders need to evaluate their tools on benchmark programs, which are currently lacking for WebAssembly.
Two datasets of WebAssembly binaries exist~\cite{Musch2019,Hilbig2021}, but they only include the binary files without the source files of the programs.
Linking a WebAssembly binary to the source code that produced it is of high value for developers of such tools, enabling them to understand the inner workings of these programs, or for example to instrument programs during their compilation.
To fill this gap, we present a methodology for identifying open-source C and C++ projects that target WebAssembly for compilation, and we build a novel dataset of WebAssembly binary files that are associated with their respective source projects.
We deploy Wasmizer, a tool that automates this process and regularly updates our dataset.

To demonstrate an application of this dataset, we investigate \emph{compilation smells} in WebAssembly, i.e., indication of potentially different behaviour from source to target programs that may yield bugs~\cite{stievenart2022}.
In fact, previous work has shown that compilers and standard library implementations available for WebAssembly are yet not as mature as those used for native compilation, and consequently, certain code patterns may yield different behaviour in these platforms~\cite{stievenart2022, Stievenart2021CompilerProtection}.
These different behaviour may yield unexpected bugs lurking around when porting programs to WebAssembly, and more importantly, they may introduce security risks, for example enabling an easier exploitation of binaries suffering from buffer overflows~\cite{Stievenart2021CompilerProtection}.
While previous work has shown these different behaviours exist for programs compiled from C and C++, we  take the next step and investigate how prevalent they are in real-world projects.

The contributions of this work are centred around the following two research questions:

\begin{itemize}
    \item \textbf{RQ$_1$: Can we mine WebAssembly-driven projects on GitHub and generate WebAssembly binaries automatically?}
    We present heuristics that help to identify real-world C and C++ projects that are related to WebAssembly. 
    We collect 2\,540 projects that target WebAssembly as a compilation target.
    We build these projects that contain a makefile and generate 8\,915 binaries from 572 repositories, forming a novel dataset of WebAssembly binaries linked to their originating projects.
    We develop Wasmizer, an open-source tool that automates this process.\footnote{\url{https://github.com/arash-mazidi/WASMIZER}}

    \item \textbf{RQ$_2$: How prevalent are compilation smells in open-source projects that target WebAssembly?}
    We present the code patterns of 16 compilation smells and heuristics that help to identify them in C and C++ programs.
    We use the Clang static analyser to develop checkers that can detect eight of such smells.
    We evaluate their presence in 1\,605 projects, uncovering that 386 projects (i.e. 24\%) suffer from at least one compilation smell.
    This analysis shows a use case of our dataset that requires access to the source code of WebAssembly applications.
    
\end{itemize}

We share our dataset, tools, and code analysis scripts to allow for replication of this work.\footnote{\url{https://doi.org/10.5281/zenodo.7742004}}
Importantly, Wasmizer is deployed to regularly mine WebAssembly-driven projects on GitHub, compile them, and curate an up-to-date dataset of WebAssembly sources and binaries.
We also share this novel and evolving dataset to facilitate further studies in this domain.\footnote{\url{http://tiny.cc/WasmizerOutput}, password: wasmizer}

\newpage

The remainder of this paper is organised as follows.
In Section~\ref{sec:background}, we provide some background on model of WebAssembly.
We present the datasets of source projects and associated binaries in Section~\ref{sec:dataset}.
In Section~\ref{sec:smells}, we study compilation smells as a use case for this dataset.
We discuss threats to validity of this work in Section~\ref{sec:threats}.
We put this work in context with the related research on WebAssembly in Section~\ref{sec:relatedwork}, and conclude this paper in Section~\ref{sec:conclusion}. %
\section{Background on WebAssembly}\label{sec:background}
We first provide background knowledge on the compilation of programs to WebAssembly, required for Section~\ref{sec:dataset}, and we explain the inner workings of WebAssembly programs, required to understand the compilation smells discussed in Section~\ref{sec:smells}.
WebAssembly is a low-level language initially aimed at bringing near-native performance for the web. It uses a stack-based execution model that runs within a virtual machine (VM) similar to the JVM. WebAssembly binaries, called modules, can be expressed in a binary or textual format. The intent is to use WebAssembly as a compile target from another source language such as C, C++, Rust C\#, Go or AssemblyScript using a compiler such as Emscripten \cite{Zakai2011}. WebAssembly modules can import and export definitions for functions and variables. These modules can be imported into JavaScript code or in other host environment and then instantiated. Each module runs within its own isolated context with its own memory and execution stack.

\subsection{Compiling from C and C++ to WebAssembly}
Multiple compilers have backends for compiling to WebAssembly nowadays. For C and C++ specifically, there are two main compiler backends: Emscripten~\cite{Zakai2011} being the original WebAssembly compiler, and Cheerp compiler~\cite{cheerpwebsite} being a competitor with a strong focus on binary size and speed.
Porting an existing C or C++ application to WebAssembly can be as simple as wrapping their build scripts with Emscripten's wrappers for \texttt{cmake} and \texttt{make}. However, the use of libraries and graphical toolkits can render porting applications more challenging.
\subsection{Memory Model}
The memory of a WebAssembly program is a linear array of bytes, called linear memory. When a module is instantiated, the memory is created with an initial size, however this can grow dynamically as needed. 
Addresses in linear memory are represented as an offset from the start of linear memory. Due to this, many common memory protection techniques such as data execution prevention and stack smashing protection are no longer needed. While this is good for the security of the host (the browser), it then falls on the developer to ensure that the code accessing this memory is bug-free, as anomalies such as buffer overflows will not be checked as long as they remain within the bounds of the linear memory. By isolating the memory used by each module, many independent instances can exist with their own memory within the same process.
\subsection{Control flow}
Unlike native code, WebAssembly only allows for structured control flow, which provides a few advantages. Firstly, it means that it is not possible to jump to arbitrary addresses, which eliminates the possibility for attacks that manipulate the control flow of the program. Instructions are placed in functions which are organised into blocks, where branches may only jump to the end of other blocks inside the current function. This control flow also allows for WebAssembly code to be validated, compiled, or transformed with only a single pass over the program.
\section{Dataset Construction}\label{sec:dataset}

We collect real-world WebAssembly projects that compile to WebAssembly.
We explain our methodology and the obtained results in this section.

\subsection{Methodology}

We use GitHub as our data source for identifying WebAssembly projects.

\subsubsection{Initial project selection}
We query the GitHub Search API to find C and C++ projects that include a few WebAssembly-related keywords.\footnote{https://docs.github.com/en/rest/search}

\subsubsection{NLP Filtering}
We apply Natural Language Processing (NLP) tools to remove projects for which WebAssembly is not an important word in the repositories' READMEs and description.

\subsubsection{Heuristic Development}
We study a number of known projects that compile to WebAssembly to familiarise ourselves with 
how WebAssembly compilers work, and to find possible ways to use WebAssembly in C/C++ projects.
We develop initial heuristics based on our observations to find specific indicators of projects that target WebAssembly.
We apply these heuristics to the projects that pass the NLP filter, and manually inspect a subset of excluded projects to expand our heuristics that eliminate projects that likely do not target WebAssembly for compilation.

\subsubsection{Building Dataset}
We obtain a first dataset of real-world projects that likely compile to WebAssembly.
We clone each repository locally.
The projects are written in C or C++, for which there is no standard build mechanism to rely on in order to build all projects.
Through a manual inspection, we notice that many projects rely either on makefiles, or on \texttt{cmake}.
Therefore, for each project:
\begin{itemize}
    \item For each \texttt{CMakeLists.txt} file, indicating the use of the \texttt{cmake} build system, we run Emscripten's \texttt{cmake} wrapper (\texttt{emcmake}).
    \item For each \texttt{Makefile}, either resulting from the previous step, or standalone, we run Emscripten's \texttt{make} wrapper (\texttt{emmake}).
\end{itemize}
We rely on Emscripten version 3.1.26, the latest at the time of writing.
We leave trying out other compilers and build script mechanisms for future work.

After applying this compilation step to all projects, we look for files that are named with either a \texttt{.wasm} or a \texttt{.wat} extension, indicating that they are WebAssembly files.
We try to convert all \texttt{.wat} files found into their binary version (\texttt{.wasm}) relying on the \texttt{wat2wasm} tool (version 1.0.31, the latest at the time of writing) from the WebAssembly Binary Toolkit,\footnote{\url{https://github.com/WebAssembly/wabt}} configured with the \texttt{--enable-all} flag to enable all available WebAssembly extensions.
We then store all \texttt{.wasm} files in a new directory, under a name composed of the SHA-256 sum of their content. This is to ensure that duplicate files are only present once in the dataset.

\subsection{Results}

\subsubsection{Initial Projects}
We use the GitHub search API to search for projects that include keywords listed in Table~\ref{table:keywords}.
We search in the repository’s title, description, README, or topics (keywords that can be added to a project).
We exclude forks as well as any project that is not updated since the official release of WebAssembly specification (December 2019).
We search projects refining by date periods that span over less than 1000 projects in order to overcome GitHub's limit of 1000 projects per query.
To account for GitHub's query rate limit, we wait for 20 seconds between two requests.
We gather a total of 6\,095 projects that list C or C++ as their source language.

\begin{table}[h!]
\centering
\caption{Wasm-related keywords used for initial query.}
\label{table:keywords}
\begin{tabular}{ll} 
 \toprule
 \textbf{Keyword}\footnote{The GitHub Search API queries are case-insensitive, so each keyword includes all differently capitalized variants.} & \textbf{Explanation}\\
 \midrule
 wasm & Abbreviation of WebAssembly\\
 webassembly	& Self-explanatory\\
 web assembly & Mis-spelling of WebAssembly\\
 emscripten & The most popular compiler for C/C++ to WebAssembly\\
    \bottomrule
\end{tabular}
\end{table}

\subsubsection{NLP Filtering}

Projects may contain words related to WebAssembly without the aim of targeting it for compilation. In order to eliminate unrelated projects, we use NLP tools to estimate the importance of WebAssembly-related keywords in each project’s README and description.
If an NLP tool detects a word as important to the text, it indicates that the project may be directly related to the term rather than a passing word. We implement the NLP step by applying the TopicRank algorithm, an enhanced variation of the PageRank algorithm provided by a Python NLP library named PyTextRank~\cite{Bougouin2013}. The algorithm decides the ranking of words by a ``voting'' or ``recommendation'' system. A lemma graph is produced with vertices representing each word in the README or description of the project selected.
These vertices contain edges that represent the voting system, where more edges mean a higher rating. Using PageRank, ranks are calculated for each vertex and returned.

While we aim to filter our dataset, we still want to be inclusive of possible relevant projects and ensure the NLP algorithm implementation is able to reliably detect keywords by comparing the results to those produced by other popular keyword extraction libraries.
We therefore run TopicRank against other popular NLP tools, namely Yake~\cite{LIAAD2021} and spaCy~\cite{spacywebsite}. 
We note that TopicRank produces similar results to the alternative tools, but also it is the most inclusive one as it detects more keywords, with the instances of the words in Table~\ref{table:keywords} being more prevalent in our results. Using the TopicRank algorithm, we detect 4\,022 projects that emphasise the keywords chosen to create our filtered dataset.

To ensure that we are not filtering out relevant projects in this step, we manually analyse 10\% of the excluded repositories.
We look for indications of compilation to WebAssembly. Most projects have one of three traits.
\begin{enumerate}
    \item They are either small test projects with very few files where there would be no instances of a README or description. These projects may compile to WebAssembly but are of lower quality and likely to be toy projects, so it makes sense to exclude them.
    \item There would be instances of users trying to port C libraries to run with WebAssembly. Most of these projects are abandoned or unfinished, denoting low quality. 
    \item The projects do not compile to WebAssembly at all.
\end{enumerate}
Based on this manual analysis, we determine that the NLP step indeed removes lower-quality and unrelated projects from our dataset while keeping the relevant projects.

\subsubsection{Heuristics for Compilation to WebAssembly}

Our dataset, after NLP filtering, contains repositories that indicate they are C or C++ projects related to WebAssembly.
However, we cannot guarantee that all the projects will compile to WebAssembly. 
We develop heuristics to filter our dataset more strictly into projects that will compile to WebAssembly.

We first study projects that are mentioned as top WebAssembly projects to learn about compiler calls and WebAssembly-related headers.
These projects are listed on LibHunt,\footnote{https://www.libhunt.com/topic/webassembly}
Awesome Open Source,\footnote{https://awesomeopensource.com/projects/webassembly}
and on the Emscripten Wiki page.\footnote{https://github.com/emscripten-core/emscripten/wiki/Porting-Examples-and-Demos}
We develop initial heuristics based on what we learn from these projects and look for them within source code from our NLP-filtered projects. Through this process, we identify 1\,499 projects that would compile into WebAssembly.
To verify our heuristics, we manually analyze 141 projects undetected by our initial heuristics, and we create refined heuristics that more strongly indicate projects that compile to WebAssembly.

Of the 141 repositories analysed, 63 do not aim to compile to WebAssembly or have not finished implementing this functionality. In the case where the projects do not aim for compilation, some instead mention unimplemented WebAssembly-related libraries: for example, the \texttt{2log.io} project mentions the QPA WASM plugin from Qt, however, it does not implement it.\footnote{https://github.com/2log-io/2log.io}
Of the case where functionality has not been implemented yet, most of the projects aim to compile to WebAssembly however, have it as a ``todo'' in their READMEs or descriptions. The intention to compile to WebAssembly indicates that we may be able to analyse it. However, we cannot judge whether all implementations have been complete in their C source code and do not add them to our dataset. 

Another challenge is that multiple repositories create their own custom WebAssembly module or require the user to manually set environment variables that are necessary for compiler calls. In our manual analysis, we could not detect common compilation indications from their code.

Based on the projects that do target WebAssembly for compilation, we create three general heuristics that we discuss below.

\paragraph{Reference to a WebAssembly compiler in the build script of the project} While not all repositories contain build scripts, the presence of compilation calls to WebAssembly within this set of projects indicates that it does target WebAssembly for compilation. We look for the presence of calls to three WebAssembly compilers for C/C++, being \texttt{emcc} (Emscripten)~\cite{Zakai2011}, \texttt{-target cheerp-wasm} (Cheerp)~\cite{cheerpwebsite} and \texttt{--target=wasm32} (LLVM)\footnote{While Emscripten also uses LLVM, this refers to projects compiled with LLVM without Emscripten.}~\cite{llvmwebsite}.
\paragraph{Inclusion of WebAssembly specific libraries} We look for mention of headers for the Emscripten APIs, as these indicate that the source code is interacting with web APIs. We focus specifically on projects that contain \texttt{\#include} macros to include the \texttt{emscripten.h} or \texttt{html5.h} headers, as these have distinct names that are easy to detect, and also provide the most commonly used core functionality.
\paragraph{Use of the JavaScript WebAssembly API} We look for instantiation of the \texttt{WebAssembly} class within JavaScript code in the project. While this on its own does not indicate compilation to WebAssembly, we found that inclusion of this code in a C/C++ project was a good indication that it compiled to WebAssembly.

\begin{table}[h!]
\centering
\caption{Manual analysis of compilation methods}
\label{table:2}
\begin{tabular}{m{6cm}  l} 
 \toprule
 \textbf{Category} & \textbf{Projects} \\
 \midrule
 Emscripten header undetected & 12 \\ 
 Does not aim to compile to WebAssembly & 31 \\
 Unfinished implementation & 32 \\
 Uses JavaScript to compile & 28 \\
 Uses C source code to compile & 9 \\ 
 Uses unique compiler methods & 8 \\
 Uses custom variables & 16 \\
 Undefined Wasm implementation & 5 \\
 \bottomrule
\end{tabular}
\end{table}

\subsubsection{Dataset}
We collect 2\,540 C/C++ projects that are related to WebAssembly i.e., include evidence that they compile to WebAssembly. 
In this dataset, 63\% of the projects have at least one star, 27\% have at least one open issue, and 41\% have been forked at least once. The average size of a repository in our dataset is 14MB. On average, the projects are over three years old, and of the 85\% of projects that are more than a year old, 57\% were still being updated a year later. 74\% of repositories list C++ as a source language, while 73\% list C as a source language. Additionally, 56\% and 49\% of projects contain HTML and JavaScript respectively.

We clone and compile each WebAssembly-related project.
This results in 8\,915 WebAssembly binaries (\texttt{.wasm} files), extracted from 572 repositories.\footnote{During the conversion from textual representation to the binary representation, 1\,384  \texttt{.wat} files could not be converted due to features unsupported by \texttt{wat2wasm} or syntax errors. We do not count these files and include them in a separate folder of our dataset.} 
None of these files overlap with the dataset of Hilbig et al.~\cite{Hilbig2021} built in 2020, indicating that all files in our dataset are new benchmark programs.

We notice that many of the \texttt{.wasm} files are already present in the repositories \emph{before} the compilation. In total, the compilation step produces 1\,096 new binary files, for 124 repositories. These binary files can be linked to their source code. The link between C and C++ source projects and a WebAssembly binary has not been part of any previous dataset so far~\cite{Musch2019,Hilbig2021}.

\begin{figure*}[h]
\centerline{
\includegraphics[height=7cm]{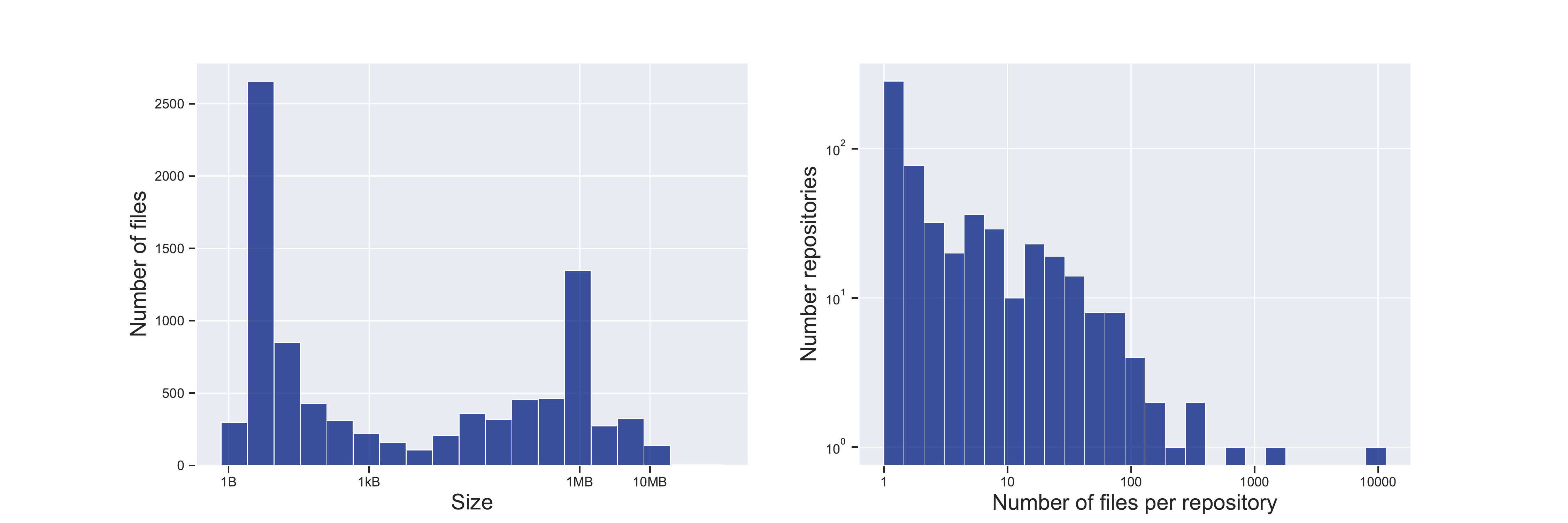}
}
\caption{Distribution of binary WebAssembly files.}\label{fig:statistics-binaries}
\end{figure*}

The distribution of the size of the binary files is represented in the left-hand side of Figure~\ref{fig:statistics-binaries}. We notice a bimodal distribution, with many files of a few bytes, and many files of around 1MB.

We find that some files are present in multiple repositories: in total, 68 files ($<$1\%) are in more than one repository, most often in two repositories, with three files being present in seven repositories.
Looking at how many WebAssembly files there are per repositories however, we notice that around half (284) of the repositories contain one file, with the other half containing more files. This distribution is represented in the right-hand side of Figure~\ref{fig:statistics-binaries}. The most extreme case is one repository containing more 11\,599 files (many of which are duplicate from each other). All but 12 repositories contain less than 100 WebAssembly binaries.

\subsection{Wasmizer}

\begin{figure*}[h]
\centerline{
\includegraphics[scale=0.7]{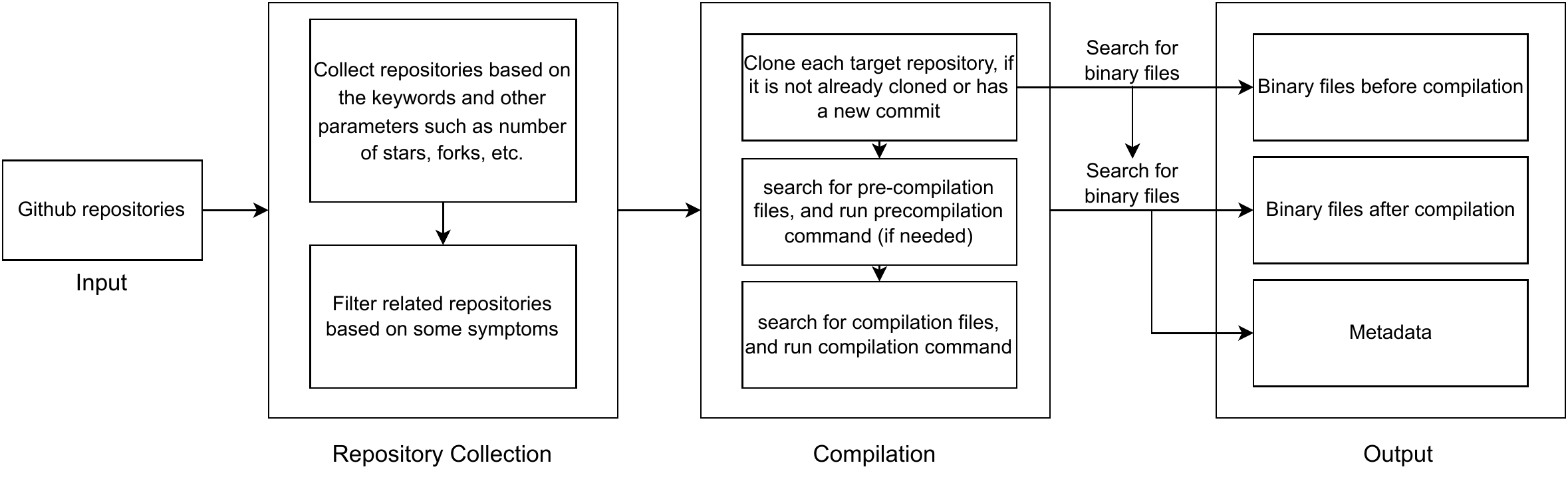}
}
\caption{The pipeline of Wasmizer.}\label{fig:wasmizer}
\end{figure*}

We develop a tool named Wasmizer that automates the entire process from project selection on GitHub to project compilation and WebAssembly binary generation. Figure~\ref{fig:wasmizer} illustrates the Wasmizer's pipeline.
Wasmizer allows for many configurable options.
It is possible to configure search parameters such as number of stars, number of forks, minimal project size, and last push date.
The filtering step, which looks for symptoms of WebAssembly compilation is also parametric.
The compilation pipeline is dynamic too: it is possible to change the compilation commands so that Wasmizer can curate programs written in other languages such as Rust or Go.

Wasmizer is open source,\footnote{\url{https://github.com/arash-mazidi/WASMIZER}} and it is deployed to regularly collect WebAssembly-driven projects on GitHub and curate an up-
to-date dataset of WebAssembly sources and binaries. 
In addition to the earlier dataset that we presented,\footnote{\url{https://doi.org/10.5281/zenodo.7742004}} we also share the ``evolving" dataset that Wasmizer is curating as of March 10th, 2023.\footnote{\url{http://tiny.cc/WasmizerOutput}}

\section{WebAssembly Compilation Smells in the Wild}\label{sec:smells}
We present a case study or our dataset by detecting the presence of compilation smells using custom checkers. We rely on existing checkers and develop new checkers for Clang to analyse our project dataset for occurrences of eight compilation smells.
Unlike previous work~\cite{stievenart2022}, which has detected these smells on synthetic code examples, we detect the smells in real source code.
In addition, we investigate more compilation configuration flags and present the code patterns that indicate these smells. 

\subsection{Compilation Smells}

\begin{figure*}[h]
\centerline{
\includegraphics[scale=0.7]{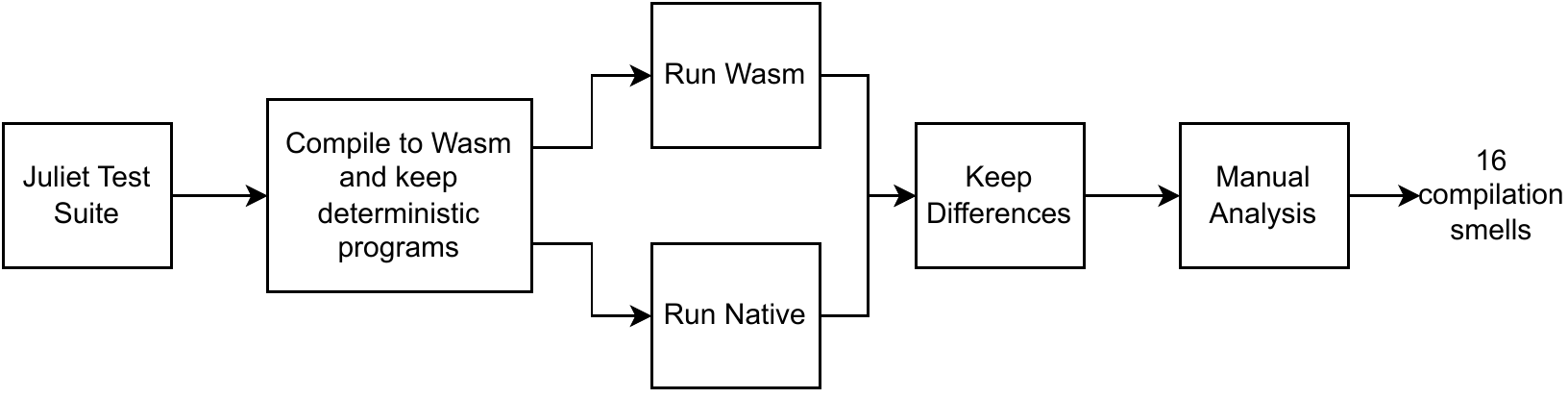}
}
\caption{The pipeline for identifying  compilation smells.}\label{fig:methodology-smells}
\end{figure*}

Compilation smells are indicators of execution differences that can arise between a C program compiled to native x86 code and the same program compiled to WebAssembly. We rely on the same dataset as the original paper~\cite{stievenart2022}, namely the Juliet test suite~\cite{boland2012juliet}, a set of test C and C++ programs that have been designed to exercise and evaluate static application security testing (SAST) tools.
We extend the previous work by repeating the experiments with multiple sets of compiler configuration flags.
This is done to ensure that the smells we identify are exhibited in various compiler configurations.
Moreover, we perform a manual analysis step to extract the pattern of each smell, which was lacking in previous work. We illustrate our pipeline to identify the smells in Figure~\ref{fig:methodology-smells}.

The Juliet test suite is composed of test programs that each has two variants: the \emph{bad} variant which contains a security issue, and the \emph{good} variant which does not contain the issue. Starting from the dataset of deterministic programs in the Juliet test suite, we compile each program to x86 and to WebAssembly. We record their execution outcome (success or crash), as well as their standard outputs. If any difference is encountered, we flag the program as behaving differently when executed in WebAssembly.

Unlike previous work, we repeat this process for a number of different compiler configurations.
In particular, we investigate the following tweaks to the default compiler configurations:\footnote{Clang flags and their default values:  \url{https://clang.llvm.org/docs/ClangCommandLineReference.html}}

\begin{itemize}
    \item The optimisation level: \texttt{-O0} (no optimisation), \texttt{-O1} (only some optimisations), \texttt{-O2} (moderate level of optimisations), and \texttt{-Os} (optimisations for code size).
    
    \item Security protections: \texttt{default}, \texttt{nonsecure} (disable a set of security protections enabled by default, and \texttt{secure} (enable a set of security protections). The flags used for each of these setting are detailed in Table~\ref{tab:clang-settings}.
\end{itemize}

\begin{table*}[ht]
\centering
\caption{Compilation settings used in our study.}
\label{tab:clang-settings}
\begin{tabular}{ll} 
 \toprule
 \textbf{Setting} & \textbf{Description} \\ \midrule
 \textbf{default} &  the default values provided by Clang 13.0.1 \\
 \textbf{nonsecure} & \texttt{-D\_FORTIFY\_SOURCE=0 -fno-pie -no-pie -fno-stack-protector}\\
 & \texttt{-fno-sanitize=safe-stack,address,undefined,cfi -fno-lto} \\
 \textbf{secure} (Wasm) & \texttt{-D\_FORTIFY\_SOURCE=2 -fpie -pie -fsanitize=cfi -flto -Wl,-z,relro}\\
 &\texttt{-Wl,-z-now -fvisibility=hidden} \\
 \textbf{secure} (x86) & \texttt{-D\_FORTIFY\_SOURCE=2 -fpie -pie -fsanitize=cfi -flto -Wl,-z,relro}\\
 &\texttt{-Wl,-z-now -fvisibility=hidden -fstack-protector}\\
 &\texttt{-fsanitize=safe-stack,cfi -flto -fuse-ld=gold}\\
 \bottomrule
\end{tabular}
\end{table*}

Moreover, unlike previous research, we target 32-bit native code instead of 64-bit. This is to ensure better compatibility with WebAssembly, which itself is a 32-bit architecture and therefore have different number semantics~\cite{Stievenart2021CompilerProtection}.

We consider 12 configurations in total :
\begin{enumerate}
\item \texttt{default,-O0,good},
\item\texttt{default,-O1,good},
\item\texttt{default,-O2,good},
\item\texttt{default,-Os,good},
\item\texttt{secure,-O2,good},
\item\texttt{nonsecure,-O2,good},
\item\texttt{default,-O0,bad},
\item\texttt{default,-O1,bad},
\item\texttt{default,-O2,bad},
\item\texttt{default,-Os,bad},
\item\texttt{secure,-O2,bad}, and
\item\texttt{nonsecure,-O2,bad}.
\end{enumerate}

Each test case is used to produce 24 binaries (12 WebAssembly binaries and 12 native binaries, each one using a different set of compiler flags).

In order to understand where these differences arise from, we manually go through each program behaving differently across multiple configurations. The structure of the Juliet test suite is so that there are many variants of the same programs, including control- and data-flow variants, e.g., by wrapping the code of the program in a specific branch that will always be taken at execution time. This allows us to reduce the set of programs to investigate manually.
The manual process includes looking at the program, hypothesizing the reason for the different behaviour, and trying to confirm that reason through other examples.

We refine the previous study~\cite{stievenart2022} that merely described the various patterns at a high-level. In this work, we manually minimize each program exhibiting a difference\footnote{We could adopt an automated approach, for example by relying on automatic program reduction techniques such as delta debugging. However, the size of the dataset and the programs within it enable us to reduce the programs manually.}: we remove portions of the code until no code can be removed without exhibiting the difference. This process results in a minimal program for each difference, which we call \emph{compilation smell}.

We find a total of 16 root causes for the differences, i.e., smells. 
We describe these smells, linking them to a CWE when relevant. We also provide insights on how each smell can be detected by automated program analysis tools.
We distinguish between \emph{structural} checks that can be performed simply by walking the AST of the program and \emph{semantic} checks that require a deeper analysis of the program semantics.
Each of these smells may have consequences when porting C and C++ programs to WebAssembly: a smell is an indication of potentially different behaviour, that itself may result in bugs (or increased security risks) in the program~\cite{Stievenart2021CompilerProtection}.

\paragraph{Double Free (CWE 415)}
The following excerpt allocates a memory region but frees it twice.
In its native version, this program yields a double free error (\texttt{free(): double free detected in tcache 2}), but it runs successfully in WebAssembly, potentially continuing the program's execution while the heap memory has been corrupted.
Instead, a \texttt{malloc}ed memory region should only be freed once.

\begin{tcolorbox}[arc=0pt,outer arc=0pt]
\begin{minted}{c}
char *data =
  (char *)malloc(100*sizeof(char));
free(data);
free(data);
\end{minted}
\end{tcolorbox}

\begin{tcolorbox}[colback=white, arc=0pt,outer arc=0pt]
Semantic check: track allocated and freed buffers to detect when \texttt{free} is called on a buffer already freed.
Already implemented in Clang as the \texttt{unix.Malloc} checker.
\end{tcolorbox}

\paragraph{Error Without Action (CWE 390)}
A file is opened with the \texttt{fopen} function.
The file is then closed, without checking that opening the file actually succeeded.
This results in an error when executed in WebAssembly (\texttt{RuntimeError: uninitialized element}), but succeeds in its native version.
The WebAssembly crash is unexpected, as the file should be created successfully after calling \texttt{fopen}.
A closer inspection reveals that \texttt{fopen} indeed fails and sets \texttt{errno} to \texttt{ENOTUNIQ} (\emph{Name not unique on network}), which is an unconventional error for \texttt{fopen}.
Instead, the return value of \texttt{fopen} should be checked before using the file handle.

\begin{tcolorbox}[arc=0pt,outer arc=0pt]
\begin{minted}{c}
FILE *f = fopen("file.txt", "w+");
fclose(f);
\end{minted}
\end{tcolorbox}

\begin{tcolorbox}[colback=white, arc=0pt,outer arc=0pt]
Semantic check: find calls to \texttt{fclose} on a file pointer that has been open with \texttt{fopen}, in a branch where it could be \texttt{NULL}.
\end{tcolorbox}

\paragraph{Double \texttt{fclose} (CWE 675)}
A file is opened, and thereafter closed twice.
In its native version, this program succeeds. In its WebAssembly version, it fails: the file is not opened successfully, resulting in the same error as we encountered in CWE 390 (\texttt{ENOTUNIQ}: \emph{Name not unique on network}).

\begin{tcolorbox}[arc=0pt,outer arc=0pt]
\begin{minted}{c}
FILE *data = 
  freopen("f.txt", "w+", stdin);
fclose(data);
fclose(data);
\end{minted}
\end{tcolorbox}

\begin{tcolorbox}[colback=white, arc=0pt,outer arc=0pt]
Semantic check: same as for "Error Without Action"
\end{tcolorbox}

\paragraph{Use of Uninitialized Variable (CWE 457)}
Similarly, the following program results in undefined behaviour because it reads data from an uninitialized behaviour. The WebAssembly output is the empty string, while the native output is the name of the program.
Instead, one should not use uninitialized variables.

\begin{tcolorbox}[arc=0pt,outer arc=0pt]
\begin{minted}{c}
char *data;
printf("%
\end{minted}
\end{tcolorbox}

\begin{tcolorbox}[colback=white, arc=0pt,outer arc=0pt]
Semantic check: find usages of variables that have not been initialized.
\end{tcolorbox}

\paragraph{Access to Environment Variables}
The following excerpt accesses the \texttt{PATH} environment variable.

\begin{tcolorbox}[arc=0pt,outer arc=0pt]
\begin{minted}{c}
printf("%
\end{minted}
\end{tcolorbox}
\begin{tcolorbox}[colback=white,arc=0pt,outer arc=0pt]
Structural check: find calls to \texttt{getenv}.
\end{tcolorbox}

In native code, this accesses a variable that is bound on Unix-like systems.
However, in WebAssembly, the execution environment is cleared upon execution of a program.
With the platform we are using to execute the test cases, it is required to explicitly bind these environment variables before executing this program.
In case this is not performed, the output of this excerpt differs between native code and WebAssembly (where the empty string is printed).
If environment variables are used in a project, they should be setup properly when instantiating the WebAssembly module.

\paragraph{Incorrect Check of Function Return Value, with \texttt{fputs} (CWE 235)}
The following example illustrates one difference that arises due to the use of \texttt{musl} as the standard C library when compiling to WebAssembly.
Function \texttt{fputs} is called, and its result is checked against the constant 0 to see if printing the string has failed.
However, the specification of \texttt{fputs} states that it returns \texttt{EOF} (-1) upon error, and a non-negative number upon success.
The \texttt{musl} library returns 0 as number upon success, while \texttt{glibc} returns the number of bytes printed.
As a result, the WebAssembly version enters the branch and prints \texttt{fputs failed!}.
Instead, one should only rely on whether the return value of \texttt{fputs} is a positive integer or not.

\begin{tcolorbox}[arc=0pt,outer arc=0pt]
\begin{minted}{c}
if (fputs("string", stdout) == 0)
    printf("fputs failed!\n");
\end{minted}
\end{tcolorbox}

\begin{tcolorbox}[colback=white,arc=0pt,outer arc=0pt]
Semantic check: find return value of \texttt{fputs} that flows into a comparison against constant 0.
\end{tcolorbox}

\paragraph{Improper Resource Shutdown (CWE404)}
A file is opened with the \texttt{open} system call.
This returns a file descriptor as an \texttt{int}.
The \texttt{fclose} function is used to close the file.
However, a file descriptor must instead be closed with the \texttt{close} system call, while \texttt{fclose} expects a pointer to a \texttt{FILE} data structure.
In WebAssembly, this code executes successfully and the program runs to completion, while it crashes in its native version.

\begin{tcolorbox}[arc=0pt,outer arc=0pt]
\begin{minted}{c}
int f = open("file.txt",
             O_RDWR | O_CREAT,
             S_IREAD | S_IWRITE);
fclose((FILE *)f);
\end{minted}
\end{tcolorbox}

\begin{tcolorbox}[colback=white,arc=0pt,outer arc=0pt]
Semantic check: find calls to \texttt{fclose} on values of type \texttt{int} which have been assigned by \texttt{open}.
\end{tcolorbox}

\paragraph{Wide Characters}
The following program simply writes a wide string to the console.
While in its native version, the string is indeed written to the console, the WebAssembly version does not print anything. It actually requires calling \texttt{fwide} to tell the console that wide characters will be printed. This is likely due to a difference of libc.
Instead, one should always call \texttt{fwide} before \texttt{wprintf}.

\begin{tcolorbox}[arc=0pt,outer arc=0pt]
\begin{minted}{c}
wprintf(L"%
\end{minted}
\end{tcolorbox}

\begin{tcolorbox}[colback=white,arc=0pt,outer arc=0pt]
Semantic check: find calls to \texttt{wprintf} that have not been preceded by a call to \texttt{fwide}.
\end{tcolorbox}

\textbf{NB.}
The code patterns that we present in the rest of this section require more advanced program analysis techniques.

\paragraph{Incorrect String Argument (CWE 688)}
The following program calls \texttt{printf} with \texttt{"\%s"} as a format string, but incorrectly passes an \texttt{int} as argument.
The native program fails with a segmentation fault, as it tries to read a string at an invalid memory location.
However, the WebAssembly binary succeeds, reading an empty string from the (invalid) memory location 5.
This is because the int argument is treated as a string pointer of which the first element is likely `0` due to WebAssembly initializing its linear memory with zeroes. As a result, this is interpreted as passing the empty string to \texttt{printf}.
Instead, calls to \texttt{printf} should ensure that the types of arguments match the format string.

\begin{tcolorbox}[arc=0pt,outer arc=0pt]
\begin{minted}{c}
printf("%
\end{minted}
\end{tcolorbox}

\begin{tcolorbox}[colback=white,arc=0pt,outer arc=0pt]
Structural check: find values of an invalid type being passed as argument to format string functions.
Already implemented in clang with \texttt{-Wformat}.
\end{tcolorbox}

\paragraph{Freeing Invalid Memory (CWE 590)}
The following program allocates memory on the stack using \texttt{alloca}, and tries to free it with \texttt{free}.
However, \texttt{free} should be used to free heap-allocated memory.
As a result, this program crashes in its native version.
However, it runs to completion in WebAssembly.
This could come from a different implementation of \texttt{free} in musl.
Instead, one should not \texttt{free} memory allocated on the stack.

\begin{tcolorbox}[arc=0pt,outer arc=0pt]
\begin{minted}{c}
char *data =
  (char *)alloca(100*sizeof(char));
free(data);
\end{minted}
\end{tcolorbox}

\begin{tcolorbox}[colback=white,arc=0pt,outer arc=0pt]
Semantic check: find memory region allocated with \texttt{alloca}, flowing into a call to \texttt{free}.
\end{tcolorbox}

\paragraph{Incorrect Number of Arguments (CWE 685)}
The following program calls \texttt{printf} with too few arguments, while the provided format string expects two arguments.
In WebAssembly, it works and treats the second string as null, because the next value on the stack likely points to a 0 and is treated as the empty string.
In native, it crashes as it cannot provide a value for the second string.
Instead, one should provide the number of arguments in agreement with the format string.

\begin{tcolorbox}[arc=0pt,outer arc=0pt]
\begin{minted}{c}
printf("%
\end{minted}
\end{tcolorbox}

\begin{tcolorbox}[colback=white,arc=0pt,outer arc=0pt]
Structural check: find calls to format string functions that provide too few arguments.
Already supported in clang with \texttt{-Wformat}.
\end{tcolorbox}

\paragraph{Freeing Pointer Not At Start of Allocated Region (CWE 761)}
The following program allocates heap memory with \texttt{malloc}, and then moves the allocated pointer further in the allocated region.
It then tries to free the memory by passing this incremented pointer as argument, which is invalid: \texttt{free} should be called on the pointer to the initial allocated region.
This crashes in native, but works in WebAssembly, due to the use of a different allocator in musl.
Instead, one should free memory regions at their starting pointer.

\begin{tcolorbox}[arc=0pt,outer arc=0pt]
\begin{minted}{c}
char *data =
  (char *)malloc(100*sizeof(char));
data++;
free(data);
\end{minted}
\end{tcolorbox}

\begin{tcolorbox}[colback=white,arc=0pt,outer arc=0pt]
Semantic check: this requires tracking allocated region and pointers to these region, detecting calls to \texttt{free} on a pointer that is not at the beginning of an allocated region.
\end{tcolorbox}

\paragraph{Pointer Subtraction (CWE 469)}
This program performs invalid pointer manipulation by subtracting two different pointers: the \texttt{slash} variable points to the slash in the first string, but is mistakenly used to compute the index of the slash in \texttt{string2}.
This prints a different value in WebAssembly and in its native version, due to difference in memory layouts, i.e., \texttt{string1} and \texttt{string2} do not have the same offsets in both binaries.
Instead, one should not subtract pointers.

\begin{tcolorbox}[arc=0pt,outer arc=0pt]
\begin{minted}{c}
char string1[] = "a/b";
char string2[] = "a/b";
char *slash = strchr(string1, '/');
printf("%
\end{minted}
\end{tcolorbox}

\begin{tcolorbox}[colback=white,arc=0pt,outer arc=0pt]
Structural check: find the subtract operation applied to two pointers.
\end{tcolorbox}

\paragraph{Stack-Based Buffer Overflow (CWE 121)}
The following program contains a buffer overflow due to an incorrect allocation: \texttt{alloca} is used to allocate 10 bytes, while it should be used to allocate 10 integers (hence, \texttt{sizeof(int) * 10} bytes).
When a buffer is copied into the badly allocated memory region, this overflows.
In its native version, this program crashes as the stack is detected as being smashed. The WebAssembly version runs the program successfully.
Instead, one should ensure that all memory accesses are made within bounds.

\begin{tcolorbox}[arc=0pt,outer arc=0pt]
\begin{minted}{c}
int *data = alloca(10);
for (int i = 0; i < 10; i++) {
  data[i] = 0;
}
\end{minted}
\end{tcolorbox}

\begin{tcolorbox}[colback=white,arc=0pt,outer arc=0pt]
Detecting buffer overflows statically is an entire research problem on its own.
\end{tcolorbox}

\paragraph{Buffer Overread (CWE 126)}
The following program illustrates a buffer overread: the \texttt{data} string is filled with 150 times the 'A' character. 99 of them are copied into the \texttt{dest} buffer, but no extra null-terminating character is added.
Hence, when printing the \texttt{dest} buffer, \texttt{printf} will continue printing the string until it encounters the byte 0.
In WebAssembly, since the memory is initialized with 0s, the likeliness of having a byte 0 right after the string, and we encounter this in practice: only 99 As are printed. In its native version however, the string contains random garbage after the 99 first bytes, and \texttt{printf} prints much more characters.
Instead, one should ensure that all memory accesses are made within bounds.

\begin{tcolorbox}[arc=0pt,outer arc=0pt]
\begin{minted}{c}
char data[150] =  "AAAAAAAA...";
char dest[100];
strncpy(dest, data, 99);
printf("%
\end{minted}
\end{tcolorbox}

\begin{tcolorbox}[colback=white,arc=0pt,outer arc=0pt]
Similarly, this kind of buffer overread is an entire research problem on its own.
\end{tcolorbox}

\paragraph{Undefined Behaviour (CWE 758)}
The following program has undefined behaviour according to the C standard.
It allocates a pointer, but does not initialize it.
When dereferencing this pointer to print a string, what will be printed is therefore undefined.
In native, the name of the program is being printed, while in WebAssembly, the empty string is printed.
Instead, one should not dereference uninitialized pointers.

\begin{tcolorbox}[arc=0pt,outer arc=0pt]
\begin{minted}{c}
char **pointer = 
  alloca(sizeof(char *));
printf("%
\end{minted}
\end{tcolorbox}

\begin{tcolorbox}[colback=white,arc=0pt,outer arc=0pt]
Semantic check: find dereferenced pointers that have not been initialized.
\end{tcolorbox}

\subsection{Smell Detection}
We rely on the Clang Static Analyzer (CSA) to detect the first eight code smells (\emph{a} to \emph{h}) in the projects in our dataset.\footnote{\url{https://clang-analyzer.llvm.org}}
It is the most suitable option as it fits the constraints imposed by the nature of the analysis. 
Precisely, CSA does not require the projects to be able to be compiled and can instead analyze individual source files, while still allowing for advanced static analyses such as dataflow analysis.

To account for unknown external factors such as input values and behaviour of libraries used, CSA uses symbolic execution and assigns unique symbols to unknown values. This allows for path-sensitive analysis as occurrences of symbols can be tracked through the execution of a program.

CSA can also be extended with custom checkers, which is required as the majority of the code patterns being searched for are not detected by existing analysis tools.
The custom checkers rely on a callback mechanism to subscribe to certain events. For example, the \texttt{check::PreCall} callback will be called every time the analyser comes across a function call before it analyses it. The analyser also supports inter-procedural analysis. The checkers have access to the program state at each point in the path analysis, and this state can be manipulated by the checkers to read and store arbitrary information.

\subsubsection{Selected Compilation Smells}
We choose the first eight compilation smells (i.e., a to h) described in the previous section. This selection is based on how feasible it is to implement checkers for these smells in the Clang static analyzer, using the tools available (e.g., no pointer analysis).

\subsubsection{Implementation of Custom Checkers}
CSA includes checkers that can identify three smells.
We implement five new custom checkers using the capabilities of the CSA. Table~\ref{tab:checkers} lists the checkers used in our experiments. 
The remaining smells (i.e., i to p) require deeper analyses such as analysis of pointers or memory allocations, which are difficult to determine statically.
We leave detection of these smells for future work.

We develop test cases (19 in total) for each custom checker to detect and eliminate both false positives and false negatives.
For example, to test the \texttt{BadFPutsComparison} checker, we implement a test case to make sure that our checker would still pick up cases where the result of calling \texttt{fputs} is compared to a variable which has a known value of 0. Similarly, for other checkers we develop tests based on the smells to ensure that the tool detects the patterns as we expect it.

\begin{table*}[ht]
\centering
\caption{Checkers used to find compilation smells. Checkers with * are new checkers that we developed.}
\label{tab:checkers}
\begin{tabular}{ll} 
 \toprule
 \textbf{Checker} & \textbf{Description} \\ \midrule
 \texttt{DoubleFree} & 2 calls to \texttt{free()} on same value \\ 
 \texttt{DoubleFClose} & 2 calls to \texttt{fclose()} on same value \\
 \texttt{UninitialisedVariable} & Checks if a variable has been initialised at the point where it is used \\
 \texttt{AccessEnv}* & Checks for calls to \texttt{getenv()} \\ 
 \texttt{BadFputsComparison}* & Comparison of \texttt{fputs()} return value to 0 \\
 \texttt{ErrorWithoutAction}* & Call \texttt{fclose()} on return value of \texttt{fopen()} without checking if it's null \\
 \texttt{ImproperResourceShutdown}* & Call \texttt{fclose()} on return value of \texttt{open()} \\
 \texttt{WideString}* & Checks if \texttt{fwide()} has not been called prior to calling \texttt{wprintf()} \\ 
 \bottomrule
\end{tabular}
\end{table*}

\subsubsection{Running the Checkers}

We run our checkers on 1\,605 projects, including both the projects that are WebAssembly-related, and the projects that are identified to target WebAssembly as a compilation target.
To do so, for each project, we move all C/C++ source files into a single directory using a Node.js script, and rename filenames to prevent conflicts. This is necessary as the Clang Static Analyzer is not able to handle files in nested subdirectories.
We then run the Clang Static Analyzer via command line, passing it the name of each C or C++ source file in the project.
This results in a \texttt{.plist} file containing the analysis results for each source file in the project. We use another Node.js script to post process the results and merge the many \texttt{.plist} files into a single file in JSON format that contains all of the key information about each code smell detected in that project.
This is a time consuming process, as some projects have up to 50\,000 source files.

\subsection{Results}

We uncover the presence of two compilation smells namely, \texttt{AccessEnv} and \texttt{ErrorWithoutAction}).
The checkers detect no instance of other six compilation smells.
Table~\ref{tab:analysis-results} lists the number of repositories affected by the two compilation smells.
We discover a total of 1\,873 compilation smells and note that 386 projects (i.e., 24\%) contain at least one instance of a compilation smell.
Future work is necessary to understand whether these smells are indication of potential bugs or not.
Nevertheless, 
the checkers that we use in this study are present in the artifact accompanying this paper, and developers can use them to detect and reduce the prevalence of compilation smells in practice.

\begin{table}[ht]
\centering
\caption{Results of Compilation Smells Analysis}
\label{tab:analysis-results}
\begin{tabular}{lrrr} 
 \toprule
 Checker & Occurences & Repositories Affected \\ 
 \midrule
 \texttt{AccessEnv} & 809 & 241 (15\%)\\
 \texttt{ErrorWithoutAction} & 974 & 300 (19\%) \\
 \bottomrule
\end{tabular}
\end{table}

None of the checkers already implemented in Clang find instances of those code patterns in our dataset. We theorise that this may be because the developers are already alerted to these code smells, such as through their IDE or during the build process, meaning that these errors are caught and rectified before they are committed to the repository.

\section{Threats to Validity}\label{sec:threats}
We define several heuristics to ensure a high-quality dataset, but every manual process is subject to bias. We may have exhibited bias in the methods we choose to sample our dataset. Such as the process of queries and keywords selected using the GitHub Search API, to the manual analysis of compilation indicators to WebAssembly. This can be shown in our selection of keywords defined in our NLP filtering. We define known keywords that we think are related to WebAssembly, but there might be more relevant keywords that we missed. 

The tools we use in this study are limited too. 
For instance, the return limit of the GitHub API reduced the number of projects we could gather, capping at 1000 results per query, even though we know that more projects exist. 
To overcome this, we search over periods of time that span less than 1000 projects.

We take reasonable steps to create a dataset that is inclusive of a wide range of projects, but it may not be representative of all C/C++ projects that compile to WebAssembly. This could be verified through analysis of a larger dataset. We also focus on open-source projects on GitHub, but they may be quite different in nature to closed-source projects developed by large companies or those projects found via other sources.

For the extraction of WebAssembly binaries from our dataset, we focus only on projects that are compiled using either makefiles directly, or relying on CMake as a build system. This is the closest we can have to a standard build system in C and C++. Many projects however could not be compiled: some do not use such build systems, e.g., sometimes resorting to shell scripts to issue the compilation commands; others require libraries that are missing; or have improper configuration of their build system. We extract WebAssembly files based on their name only (\texttt{.wasm} or \texttt{.wat}), but there could be other files that use a different naming scheme.

We share the dataset of binaries with information about the projects from which each binary is built. 
It is important to check license compatibility of these projects to ensure that any legal restrictions or obligations are properly adhered to.

The code patterns that we could detect are limited by the capabilities of the Clang static analysis tool being used. There is further scope to develop analysis tools that can detect the remaining code smells that are not implemented in this work. Detection of these code smells require modelling the memory and pointers used in the program, making them more difficult to detect using only static analysis.
Even though we detect many instances of compilation smells at the level of a project, such smells may be in a file that will not be compiled to WebAssembly, even if part of the project does.
Hence, further work is required to determine which specific files within a project will be compiled to WebAssembly.
Finally, we do not investigate the context in which these issues arise nor the impacts that such unwanted code patterns have on these projects.
This requires an intensive manual effort that we plan to conduct in future.

\section{Related Work}\label{sec:relatedwork}

We build the first dataset of WebAssembly sources and their binaries.
This dataset is large, and it remains up to date with the help of Wasmizer tool which is deployed.
We also investigate the prevalence of compilation smells in real-world WebAssembly programs.

Regarding the real-world usage of WebAssembly, Hilbig et al~\cite{Hilbig2021} found that 64.2\% of WebAssembly binaries are compiled from C or C++ in open access repositories. This is significant as these memory-unsafe languages are particularly vulnerable to security weaknesses when compiled to WebAssembly due to the lacking compiler protections. Additionally, many of these binaries have potential memory-related vulnerabilities, such as making use of the unmanaged stack, or using a custom memory allocator, which increase the risk of security weaknesses within the program~\cite{Hilbig2021}.
The lack of memory protection measures implemented in WebAssembly compilers leads to security weaknesses. Almost 80\% of binaries make use of the LLVM toolchain~\cite{Lattner2004}, so any security measures implemented in this toolchain would have a significant impact on the security of WebAssembly binaries.

When compiling C/C++ programs to WebAssembly, there have been observed differences in the behaviour of the resulting binaries when compared to x86 native code~\cite{Stievenart2021CompilerProtection, stievenart2022}. While they may introduce hard-to-detect bugs, in many cases these differences may be fairly harmless, such as differing outputs from print statements. However, as WebAssembly lacks common memory protections some behavioural differences related to memory allocation may lead to security weaknesses.
Three main causes were identified as being responsible for the observed differences~\cite{stievenart2022}. The first is a difference due to a different implementation of the C standard library used in the native executable and its WebAssembly counterpart. The second cause is the previously mentioned lacking memory protections of WebAssembly, where cases that would cause an exception to be thrown in native binaries instead continue to run in WebAssembly. Finally differences in the execution environment also account for some behaviour differences.

All security critical differences relate to the memory model and its protections (or lack thereof).
In a C program compiled to a native binary, a stack smashing attack would be prevented by a stack canary. However, as these are not present in WebAssembly, it leaves the application vulnerable to such an attack~\cite{Stievenart2021CompilerProtection, stievenart2022}. Similarly, a buffer underwrite that would usually be protected against via hardware protection or bounds checks remains undetected~\cite{stievenart2022}. While the sandboxed environment of WebAssembly prevents such attacks manipulating anything outside of the binary’s memory, Lehmann et al.~\cite{Lehmann2020} demonstrated how these vulnerabilities can be exploited in order to execute an XSS attack in a browser, run an arbitrary shell command in a node.js server-side environment, and write arbitrary content to a file in Wasmtime standalone runtime. While the work done by Stiévenart et al.~\cite{Stievenart2021CompilerProtection, stievenart2022} limits itself to only consider the Clang toolchain, McFadden et al.~\cite{Mcfadden2018} establishes that similar vulnerabilities are present in the Emscripten compiler toolchain.
These studies have shown the existence of code patterns that exhibit different behaviour in a syntactic test suite, however we aim to take the next step and investigate the prevalence of these code patterns in real world projects.

\section{Conclusion}
\label{sec:conclusion}

We identify 2\,540 C and C++ projects on GitHub that are related to WebAssembly and likely target WebAssembly for their compilation. 
We compile these projects to build a dataset of 8\,915 binaries that are linked to their corresponding source project.
We develop Wasmizer, a tool that fully automates this process.
Wasmizer is open source and is running on a dedicated machine to regularly mine GitHub projects and provide researchers with a novel and up-to-date dataset of WebAssembly binaries and their associated projects.

To present a use case of this dataset, we investigate the presence of eight WebAssembly \emph{compilation smells} in 1\,605 projects.
We find that 386 projects that aim to compile to WebAssembly exhibit at least one compilation smell. The most prevalent smells are calls to \texttt{getenv()} and calls to \texttt{fclose()} on the return value of \texttt{fopen()} without checking if the value is null. Our findings conclude that developers compiling native programs to WebAssembly should be aware of behavioural differences in their implementation and expected results. This is due to many repositories exhibiting similar issues, which may adversely affect development goals. 

In future, we plan to locate the exact source files that generate WebAssembly binaries. This can for example be achieved by tracing compiler calls or by inspecting dependencies in the build system.

\section*{Acknowledgment}
The feedback we received from the MSR 2023 reviewers was helpful in improving Wasmizer and making it a more useful tool for the research community. We are grateful for their input and excited to publicly share Wasmizer and the evolving dataset that it curates.

\bibliographystyle{IEEEtran}
\bibliography{References}

\end{document}